\documentclass[12pt]{article}
\usepackage{epsfig}
\usepackage{amsmath}
\usepackage{amsfonts}
\usepackage{amssymb}

\begin{document}
\pagestyle{empty}
\date{February 12, 2009}
\begin{center}
{\LARGE \bf  Chiral Symmetry Breaking\\ in Soft-Wall AdS/QCD}

\vspace{1.0cm}

{\sc Tony Gherghetta,$^{a,}$}\footnote{E-mail:  tgher@unimelb.edu.au}
{\sc Joseph I. Kapusta$^{b,}$}\footnote{E-mail:  kapusta@physics.umn.edu}
{\small and}
{\sc Thomas M. Kelley$^{b,}$}\footnote{E-mail:  kelley@physics.umn.edu}\\
\vspace{.5cm}
{\it\small {$^a$School of Physics, University of Melbourne, Victoria 3010,
Australia}}\\
{\it\small {$^b$School of Physics and Astronomy, University of Minnesota,\\
Minneapolis, MN 55455, USA}}\\
\end{center}

\vspace{1cm}
\begin{abstract}
We show how to incorporate chiral symmetry breaking in the soft-wall version of 
the AdS/QCD model by using a modified dilaton profile and a quartic term in the bulk scalar potential.  This allows one to separate the dependence on spontaneous and explicit chiral symmetry breaking. Moreover our 5D model automatically incorporates linear trajectories and non chiral-symmetry restoration for highly excited radial states.  We compare our resulting mass spectra in the scalar, vector and axial-vector sectors with the respective QCD resonances and find reasonable agreement using the known values for the pion mass and decay constant. 

\end{abstract}

\vfill
\begin{flushleft}
\end{flushleft}
\eject
\pagestyle{empty}
\setcounter{page}{1}
\setcounter{footnote}{0}
\pagestyle{plain}

\section{Introduction} 
\label{secIntro}

The AdS/CFT correspondence~\cite{Maldacena:1997re, Gubser:1998bc, Witten:1998qj, Klebanov:1999tb} provides a remarkable mathematical tool with which to understand strongly-coupled gauge theories. By using an effective dictionary that relates strongly-coupled gauge theories to higher-dimensional gravitational theories, calculations performed on the gravity side can be reinterpreted as due to nonperturbative effects on the field theory side. This allows for previously incalculable quantities to be calculated using the gravitational dual.  The theory of the strong interactions, or quantum chromodynamics (QCD), is a strongly-coupled gauge theory at low energies, and therefore a natural candidate with which to apply the gauge/gravity correspondence.  This had led to a bottom-up approach, commonly known as AdS/QCD~\cite{Erlich:2005qh, DaRold:2005zs}, which relates QCD to a five-dimensional (5D) gravity theory. This model is simple and predictive, capturing the essential features of the low-lying meson spectrum.

In an attempt to incorporate more realistic features of the excited states such as the linear Regge behavior of QCD, the AdS/QCD model can be modified to include a dilaton with a quadratic profile~\cite{Karch:2006pv}. While the linear radial spectrum is indeed produced in this soft-wall version of the AdS/QCD model, the form of chiral symmetry breaking is not QCD-like. In particular, the bulk scalar field, dual to the quark bilinear operator ${\bar q} q$, whose vacuum expectation value (VEV) is responsible for spontaneous chiral symmetry breaking, does not allow the spontaneous and explicit breaking to be independent. Moreover, chiral symmetry is restored for the highly excited states, a phenomenological feature that is not supported in the QCD mass spectrum~\cite{Shifman:2007xn}. 

In this paper, we modify the existing soft-wall version of the AdS/QCD model in order to incorporate these two phenomenological features of QCD. This is done by adding a quartic term to the bulk scalar potential and changing the dilaton profile. By assuming that the bulk scalar field contains the desired limiting behavior for non restoration of chiral symmetry \cite{Klebanov:1999tb, Shifman:2007xn}, we derive a new dilaton background profile. The extra parameter introduced by the quartic term decouples the quark mass from the chiral condensate, thereby allowing for spontaneous and explicit chiral symmetry breaking to occur independently.  The dilaton profile resulting from the required form of the bulk scalar field VEV conforms to the expected behavior required to obtain linear trajectories at large conformal coordinate $z$. The small $z$ behavior of the dilaton modifies the potential, thus affecting the mesons in the extra dimension and producing different mass spectra for the mesons under consideration.

Of course, constructing an AdS dual theory that encompasses the richness of QCD presents the greatest challenge, a task yet to be accomplished. In fact certain QCD-like characteristics such as event shapes in high energy collisions are not reproduced in the simple 5D gravitational model~ \cite{Polchinski:2002jw, Hofman:2008ar, Csaki:2008dt}. Instead genuine stringy dynamics seem to be required to capture the complete QCD behavior. Nevertheless the remarkable fact that at large t'Hooft coupling certain features of the hadron spectrum can be calculated by a relatively simple 5D model make the AdS/QCD correspondence an important analytic tool that can further our understanding of QCD.

The outline of this paper is as follows: We introduce the modified soft-wall model in Section \ref{secDual} and describe how the new dilaton profile and higher-order terms lead to a model describing spontaneous and explicit chiral symmetry breaking.  Our fit to the QCD meson spectrum is made in Section 3 where we analyze the scalar, vector, and axial-vector sectors of the model. In Section 4 we determine the pion decay constant, the Gell-Mann-Oakes-Renner relation, the coupling of the vector mesons to the pions, and the pion electromagnetic form factor.  We conclude with a discussion in Section~5.

\section{The 5D Model} 
\label{secDual}

We will consider a modified version of the soft-wall AdS/QCD model first introduced in 
\cite{Karch:2006pv} and further investigated in \cite{Evans:2006ea,Grigoryan:2007my,Kwee:2007nq,Cherman:2008eh,Colangelo:2008us,Batell:2008zm,Huang:2007fv}. 
The background geometry is assumed to be 5D AdS space with the metric
\begin{equation}
ds^{2}=g_{MN}dx^{M}dx^{N}=a^2(z)\left( \eta_{\mu\nu}dx^{\mu}dx^{\nu}+dz^{2}\right)~,
\end{equation}
where $a(z)=L/z$ is the warp factor, $L$ is the AdS curvature radius and the Minkowski metric $\eta_{\mu\nu}={\rm diag}(-1,+1,+1,+1)$. The conformal coordinate $z$ has a range $0\leq z < \infty$. To obtain linear trajectories, \cite{Karch:2006pv} also introduced a background dilaton field, $\phi$, with the asymptotic behavior
\begin{equation} \label{dilatonlz}
\phi(z\rightarrow \infty) \simeq \lambda z^2,
\end{equation}
where $\lambda$ sets the mass scale for the meson spectrum. The varying dilaton field also ensures that conformal symmetry is gradually broken in this phenomenological dual theory. 

To describe chiral symmetry breaking in the meson sector the 5D action contains
SU(2)$_L\times$ SU(2)$_R$ gauge fields and a bifundamental scalar field $X$. As suggested by \cite{Karch:2006pv}, we add a quartic term in the potential $V(X)$ of our 5D action,
\begin{equation} \label{action1}
S_{5}=-\int d^{5}x \sqrt{-g}\,e^{-\phi(z)}{\rm Tr}\left[|D X|^{2}+ m_{X}^{2} |X|^{2}-\kappa |X|^{4}+\frac{1}{4 g_{5}^{2}}(F_{L}^{2}+F_{R}^{2})\right],
\end{equation}
where $m_{X}^{2} = - 3/L^{2}$, $\kappa$ is a constant and $g_5^2=12\pi^2/N_c$, with $N_c$ the number of colors. The field tensors $F_{L,R}$ are defined as
\begin{displaymath}
F_{L,R}^{MN}=\partial^{M}{A_{L,R}^{N}}-\partial^{N}{A_{L,R}^{M}}-i[A_{L,R}^{M},A_{L,R}^{N}],
\end{displaymath} 
where $A_{L,R}^{MN}= A_{L,R}^{MNa} t^a$ and Tr$[t^at^b]=\delta^{ab}/2$, 
and the covariant derivative becomes $D^M X=\partial^M X-i A_L^MX+iXA_R^M$. To describe the vector and axial-vector mesons we simply transform to the vector (V) and axial-vector fields (A) where $V^{M}=\frac{1}{2}(A_{L}^{M}+A_{R}^{M})$ and $A^{M}=\frac{1}{2}(A_{L}^{M}-A_{R}^{M})$.

\subsection{Bulk scalar VEV solution} 
\label{secFields}

The scalar field, $X$, which is dual to the operator $\bar{q} q$, is assumed to obtain a $z$-dependent vacuum expectation value (VEV), 
\begin{equation}
\label{xvev}
\langle X \rangle \equiv \frac{v(z)}{2}
\left( \begin{array}{cc} 
  1 & 0 \\
  0 & 1  
\end{array} \right),
\end{equation}
which breaks the chiral symmetry SU(2)$_L\times$ SU(2)$_R \rightarrow$ SU(2)$_V$. Assuming (\ref{xvev}) we obtain a nonlinear equation for the VEV $v(z)$,
\begin{equation}
\partial_z(a^3 e^{-\phi} \partial_z v(z))-  a^5 e^{-\phi} (m_X^2 v(z)-\frac{\kappa}{2} v^3(z))=0.
\label{VEVequation}
\end{equation}
When $\kappa=0$, the solution of (\ref{VEVequation}) which leads to a finite action in the limit $z\rightarrow\infty$ is given 
by~\cite{Karch:2006pv, Colangelo:2008us}
\begin{equation} \label{equk0}
v(z)\simeq m_q z \, U\left(\frac{1}{2},0,z^{2}\right),
\end{equation}
where $U(a,b,y)$ is the Tricomi confluent hypergeometric function. Note that with a UV boundary located at $z=z_0$, a boundary mass term for the scalar field
needs to be added so that (\ref{equk0}) remains a consistent solution.

As expected from the AdS/CFT dictionary established in \cite{Klebanov:1999tb, Witten:1998qj}, the VEV as $z\rightarrow 0$ should take the asymptotic form
\begin{equation} \label{asymptote}
v(z)= \alpha z + \beta z^3~.
\end{equation}
The quark mass $m_q$ and the chiral condensate $\langle \bar{q} q \rangle\equiv \Sigma$, are then related to the constants in (\ref{asymptote}), via
\begin{eqnarray}
m_q&=&\frac{\alpha L}{\zeta}~, \label{quarkm} \\
\Sigma &= & \beta L \zeta~, \label{condensate}
\end{eqnarray}
where $\zeta$ is the normalization parameter introduced in \cite{Cherman:2008eh}. For fixed values of $m_{q}$ and $\Sigma$, the introduction of $\zeta$ still satisfies the Gell-Mann-Oakes-Renner relation, $m_\pi^2 f_\pi^2=2 m_q\Sigma$.
Expanding the solution (\ref{equk0}) in the small $z$ limit leads to $\alpha\propto m_q$ and $\beta\propto \Sigma \propto m_q$. Thus in the limit $m_q \rightarrow 0$ the model eliminates explicit and spontaneous chiral symmetry breaking in contradiction with QCD. It will be seen that the introduction of a quartic term in the potential $V(X)$ avoids the dependence of the chiral condensate on the quark mass encountered in \cite{Karch:2006pv,Kwee:2007nq,Colangelo:2008us}. 

Furthermore, the solution (\ref{equk0}) has an asymptotic limit $v(z)\rightarrow$ constant for large values of $z$.  This asymptotic behavior suggests chiral symmetry restoration in the mass spectrum, a phenomenon not supported in QCD (although speculation continues on whether such a restoration indeed exists \cite{Cohen:2005am,Wagenbrunn:2006cs}). As noted in \cite{Shifman:2007xn} the highly excited mesons exhibit seemingly parallel slopes signifying that chiral symmetry is not restored. In order to incorporate this behavior the scalar VEV $v(z)$ must behave linearly as $z$ becomes large,
\begin{equation}
v(z\rightarrow\infty) \sim z,
\end{equation}
causing the mass difference between vector and axial-vector resonances to approach a constant as $z\rightarrow\infty$. By including a quartic term and requiring $v(z)$ to have this linear asymptotic behavior we aim to incorporate these QCD-like characteristics into the soft-wall model.

\subsection{A new parametrized solution}

The solution for the VEV $v(z)$ was derived from the dilaton form (\ref{dilatonlz}) and, as we have seen, does not reproduce the phenomenological features expected in QCD.  Instead of solving for $v(z)$ directly, we assume the VEV asymptotically behaves as expected, namely
\begin{eqnarray}
v(z\rightarrow 0) &=& \frac{m_{q} \zeta }{L} z + \frac{\Sigma}{\zeta L} z^{3}, \label{smallV} \\
v(z\rightarrow\infty) &=& \frac{\gamma}{L} z, \label{largeV}
\end{eqnarray}
and then solve for the dilaton $\phi(z)$ using (\ref{VEVequation}) which becomes
\begin{equation}
\label{phieqn}
\phi'(z)=\frac{1}{a^3 v'(z)}\left[\partial_z(a^3 v'(z))-a^5 (m_X^2 v(z)- \frac{\kappa}{2} v^3(z))\right],
\end{equation}
where the prime $(')$ denotes the derivative with respect to $z$. Given the required behavior (\ref{smallV}) and (\ref{largeV}) we can uniquely determine the dilaton profile up to a constant. With this procedure the two sources of chiral symmetry breaking decouple while simultaneously allowing for linear trajectories in the meson spectrum.

\begin{figure}[h!]
\begin{center}
\includegraphics[scale=0.49,angle=90]{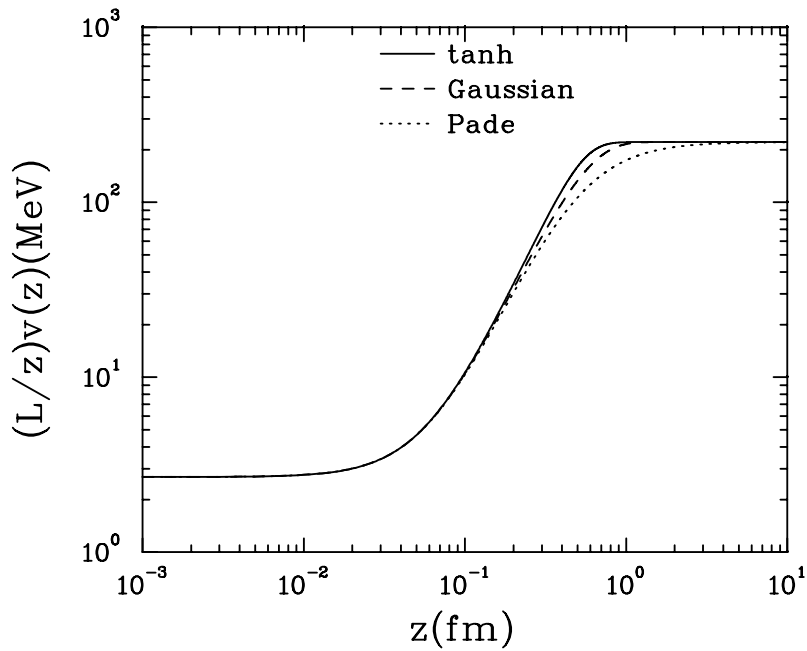}
\caption{\textsl{A plot of $v(z)/z$ for various parameterizations fitted to the mass spectra. The best fit to the mass spectra is obtained with the tanh form (\ref{arcv}).}}
\label{vevfig}
\end{center}
\end{figure}

A particularly simple parametrized form for $v(z)$ that satisfies (\ref{smallV}) and (\ref{largeV}) is
\begin{equation}
v(z) = \frac{z}{L}(A + B \tanh{C z^2}), \label{arcv}
\end{equation} 
where $A$, $B$, and $C$ are all positive coefficients dependent upon $m_{q}$, $\Sigma$, $N_{c}$, and $\kappa$, as plotted in Figure \ref{vevfig}. Expanding (\ref{arcv}) at small and large $z$ leads to the desired asymptotic forms
\begin{eqnarray}
v(z\rightarrow 0)L &=&  A z + B C z^3+{\cal O}(z^5), \label{arcvsmall} \\
v(z\rightarrow \infty)L &=&  (A+B) z. \label{arcvlarge}
\end{eqnarray}
When $A=0$, corresponding to a zero quark mass, the coefficients of the cubic term in (\ref{arcvsmall}) and of the linear term in (\ref{arcvlarge}) are nonzero, implying a nonzero chiral condensate and non restoration of chiral symmetry. Alternatively, when $B=0$ (or $C=0$), corresponding to a zero chiral condensate, the coefficients of the linear terms in (\ref{arcvsmall}) and (\ref{arcvlarge}) are both nonzero, implying a nonzero quark mass and non restoration of chiral symmetry. Thus the parametrized form in (\ref{arcv}) allows the sources of spontaneous and explicit chiral symmetry breaking to remain independent. 

Substituting (\ref{arcv}) into (\ref{phieqn}) leads to the following asymptotic behavior
\begin{eqnarray}
\phi(z\rightarrow 0) &=& \frac{\kappa}{4} A^2 z^2 + \mathcal{O}(z^6), \label{arcphi0} \\
\phi(z\rightarrow\infty) &=& \frac{\kappa}{4} (A+B)^2 z^2, \label{arcphiinf}
\end{eqnarray}
where we have chosen the integration constant arising from (\ref{phieqn}) to be zero in order for the background to be conformally invariant at $z=0$. To reproduce the limits (\ref{arcvsmall}) and (\ref{arcvlarge}) the dilaton profile at small $z$ (\ref{arcphi0}) must differ from that at large $z$ (\ref{arcphiinf}). Importantly this does not sacrifice the linear trajectories which (as will be shown) depend on the dilaton having the asymptotic form (\ref{dilatonlz}).  Note that the quartic term with strength $\kappa$ is necessary to obtain the required behavior.  Therefore, modifying the dilaton and introducing quartic interaction terms in the Lagrangian is necessary to improve the soft-wall version of the AdS/QCD model.

The normalization $\zeta$ is not a free parameter but is determined by QCD as shown in \cite{Cherman:2008eh}, namely, $\zeta=\sqrt{N_c}/(2\pi)=\sqrt{3}/g_5$. Then the parameters $\gamma$, $A$, $B$ and $C$ can be expressed in terms of the input parameters $m_q,\Sigma,\lambda,\kappa$, 
\begin{eqnarray}
\gamma &=& \sqrt{\frac{4 \lambda}{\kappa}}, \label{padeC}\\
A &=& \frac{\sqrt{3}m_q}{g_5}, \label{arcA} \\
B &=& \gamma - A, \label{arcB} \\
C &=& \frac{g_5 \Sigma}{\sqrt{3}B}. \label{arcC}
\end{eqnarray}
The input parameters are determined as follows.  The parameter $\lambda$ is determined by the average slope of the radial trajectories of the scalar, vector, and axial-vector mesons for radial quantum numbers $n \ge 3$. Its value was determined to be $\lambda = 0.1831$ GeV$^2$, as explained in the next section.  The quark mass, quark condensate, pion decay constant, and pion mass are all related by the Gell-Mann-Oakes-Renner relation, $f_{\pi}^2 m_{\pi}^2 = 2 m_q \Sigma$.  This relation holds in this model as a natural consequence of chiral symmetry \cite{Erlich:2005qh}; see also section \ref{secPion}.  We use the measured values of $f_{\pi}=92.4$ MeV and $m_{\pi}=139.6$ MeV, and adjust the quark mass to reproduce the input value of $f_{\pi}$ from a solution to the axial-vector field equation in Section~\ref{secPion} for a given value of $\kappa$.  The parameter $\kappa$ essentially controls the mass splitting between the vector and axial-vector mesons.  It is determined to be $\kappa = 15$ by a best fit to the radial spectra of the axial-vector mesons, also shown in the next section.  This results in $m_q = 9.75$ MeV and therefore $\Sigma =(204.5$ MeV)$^3$.  The inferred value of the quark mass is consistent with an average of the up and down quark masses as summarized in the Review of Particle Physics \cite{pdg} as appropriate at the hadronic energy scale.
  
Note that other parameterizations of the VEV $v(z)$ which lead to qualitatively similar behavior as that required in (\ref{smallV}) and (\ref{largeV}) can also be used. In particular, other forms for $v(z)/z$ include
$(a_1 + a_2 z^2)/(1 + a_3 z^2)$ [Pade], $b_1 + b_2 \arctan(b_3 z^2)$, $c_1 - c_2 \exp(-c_3 z^2)$ [Gaussian], $d_1 \tanh((z^2+d_3^2)/d_2^2)$, 
and $e_1 + e_2 \tanh^2(e_3 z)$. These forms were all studied but the best results were found to be obtained using the form (\ref{arcv}).  The tanh parameterization (\ref{arcv}), as well as the Pade and Gaussian forms, are shown in Figure 1 using the above parameters.  The resulting plots of $d\phi/dz$, which enters the differential equations that determine the mass spectra, and the dilaton profile $\phi(z)$, are shown in Figures 2 and 3. It becomes quite evident from the figures that a small change in $v(z)$ parametrization leads to drastic change in the behavior of the dilaton $\phi(z)$.

\begin{figure}[h!]
\begin{center}
\includegraphics[scale=0.49,angle=90]{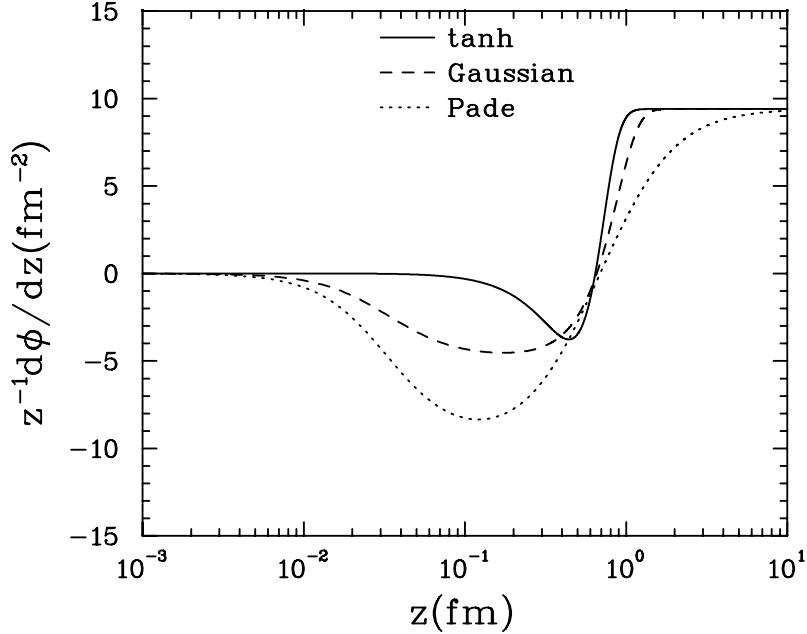}
\caption{\textsl{A plot of $\phi'(z)/z$ derived from the various parameterizations of $v(z)$. The best fit to the mass spectra is obtained with the tanh form (\ref{arcv}).
}}
\label{dotdilatonfig}
\end{center}
\end{figure}

\begin{figure}[h!]
\begin{center}
\includegraphics[scale=0.49,angle=90]{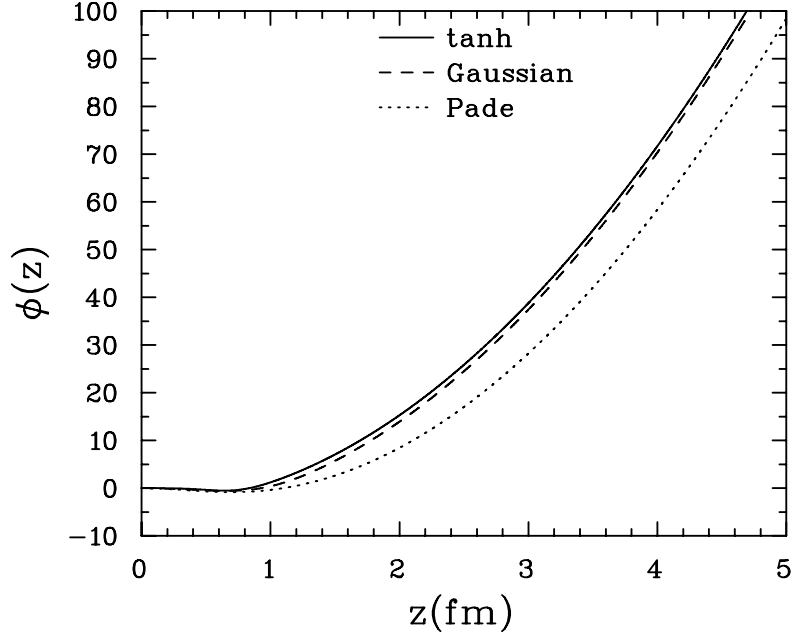}
\caption{\textsl{The dilaton profile $\phi(z)$ resulting from the various parameterizations of $v(z)$. The best fit to the
meson spectra occurs with the tanh parameterization (\ref{arcv}). For $z\lesssim1$ the behavior deviates from the quadratic asymptotic form (\ref{dilatonlz}).
}}
\label{dilatonfig}
\end{center}
\end{figure}

\section{Meson Mass Spectra}

The soft wall model can be used to fit the meson mass spectra and it is interesting to see how well this simple 5D model matches real data.  The scalar, vector and axial-vector resonances used in our fits are given in Tables 1, 2 and 3.  All but one of the included states are listed in the Review of Particle Physics (RPP)~\cite{pdg}.  A notable absence is the $\rho(1570)$ which may be an OZI violating decay of the $\rho(1700)$.  The review by Bugg \cite{Bugg} lists the state $f_0(2020 \pm 38)$ which we interpret to be the same state listed in the RPP as $f_0(1992 \pm 16)$.  The RPP lists the $f_0(2103 \pm 8)$, following Bugg's $f_0(2102 \pm 13)$, which nicely fits the $n=7$ radial excitation.  The RPP also lists the $f_0(2314 \pm 25)$ based on Bugg's $f_0(2337 \pm 14)$, which would be $n=8$.  The most uncertainty lies with the scalar mesons since mixing is expected among light quark mesons, four quark states, $s\bar{s}$ mesons, and glueballs.  This could shift the masses of the ``pure'' radial excitations of the lightest scalar meson by ${\cal O}(100$ MeV).  As pointed out in \cite{Bugg}, it has long been known that the $\rho(1465)$ is too massive to be the first radial excitation of the $\rho(775)$.  Reference \cite{Bertin} studied the reaction $p + \bar{p} \rightarrow 2\pi^+ + 2\pi^-$.  They infer the $n=2$ radial excitation of the $\rho$ to be 1282$\pm$37, which is the value used in our fits.  

\begin{figure}[h!]
\begin{center}
\includegraphics[scale=0.49,angle=90]{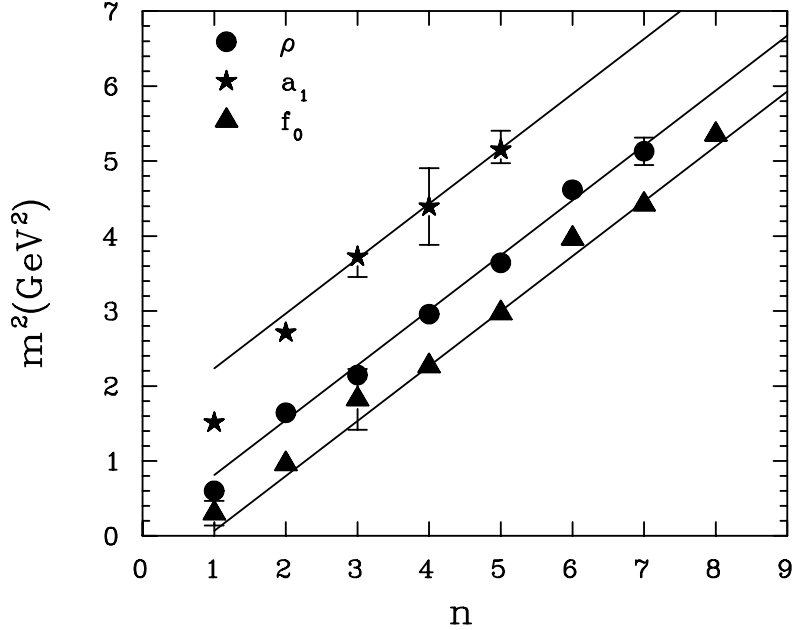}
\caption{\textsl{A straight-line fit to the measured scalar, vector and axial-vector mass spectra for $n \ge 3$ used to determine the dilaton mass parameter
$\lambda$. 
}}
\label{combinedplot}
\end{center}
\end{figure}

A straight line is fitted to the $m^2$ versus $n$ plot with $n \ge 3$ for all three mesons, assuming the same slope $4\lambda$ but different intercepts, namely $m^2_n = 4\lambda n + m^2_0$.  The results are: $\lambda = 0.1831 \pm 0.0059$ GeV$^2$, $m_{V,0}^2 = 0.0806 \pm 0.0104$ GeV$^2$, $m_{A,0}^2 = 1.5023 \pm 0.0366$ GeV$^2$, and $m_{S,0}^2 = -0.6634 \pm 0.0038$ GeV$^2$.  See Figure~\ref{combinedplot}.  We use this value of $\lambda$ in our model calculations. 

\subsection{Scalar sector} 
\label{secScalar}

Introducing a quartic term in the Lagrangian causes the scalar excitations to couple with their own VEV, giving a modified equation of motion unlike those in \cite{DaRold:2005zs, Colangelo:2008us}. Assuming $X(x,z)\equiv (v(z)/2+S(x,z))e^{2 i\pi(x,z)}$, with $\pi(x,z)$ the pion field and $S(x,z)={\cal S}_n(x) S_n(z)$, we obtain
\begin{equation}
\partial_z(a^3 e^{-\phi}\partial_zS_n(z))- a^5 e^{-\phi}(m_X^{2} -\frac{3}{2}\kappa v^2(z)) S_n(z)= -a^3 e^{-\phi} m_{S_n}^2 S_n(z), \label{scalarequ}
\end{equation}
where $S_n(z)$ are the Kaluza-Klein modes and only linear terms in $S_n$ have been kept in (\ref{scalarequ}). Note that by ignoring the nonlinear terms in (\ref{scalarequ}) we are assuming infinitesimally small amplitudes $S_n$. Because of the $z$-dependent mass term, (\ref{scalarequ}) is difficult to solve analytically for the parametrized solution of $v(z)$; however, we implement a shooting method in which (\ref{scalarequ}) is solved for various values of $m_{S_n}$. The eigenvalues are then those mass values that produce a solution for $S_{n}(z)$ that is bounded as $z\rightarrow\infty$.

\begin{figure}[h!]
\begin{center}
\includegraphics[scale=0.45,angle=90]{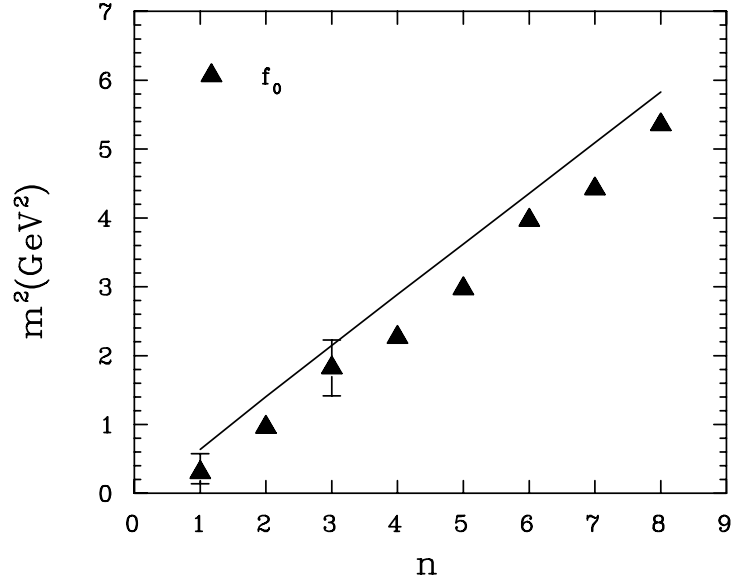}
\caption{\textsl{Comparison of the predicted scalar mass eigenvalues using the tanh form (\ref{arcv}) of $v(z)$ (solid) with the QCD $f_{0}$ scalar mass spectrum \cite{pdg}. 
}}
\label{ScalarMasses}
\end{center}
\end{figure}

The scalar equation of motion (\ref{scalarequ}) can be brought into a Schr\"{o}dinger-like form by using the substitution 
\begin{equation}
S_n(z)=e^{\omega_s/2}s_n(z),
\end{equation} 
where $\omega_s = \phi(z) + 3\log{z}$ and leads to
\begin{equation} \label{schroS}
-\partial_z^{2}s_n(z)+\left(\frac{1}{4}\omega_s'^2-\frac{1}{2}\omega_s''-\frac{3}{2}\frac{L^2}{z^2} \kappa v^2(z)-\frac{3}{z^2}\right)s_n(z)=m_{S_n}^{2}s_n(z).
\end{equation}
Applying the shooting method to (\ref{schroS}) 
with the boundary conditions $\lim_{z_0\rightarrow 0} s_n(z_0)=0$, $ \partial_z s_{n}(z\rightarrow\infty)=0$ produces the scalar mass spectra listed in Table~\ref{scalarmasses}, and displayed in Figure \ref{ScalarMasses}. The reproduction of the experimentally measured masses is reasonable, apart from an overall normalization.  This could well be a failure of this specific model.  However, considering that these light quark/antiquark radial excitations mix with scalar $s\bar{s}$ excitations, scalar glueballs, and possible four quark states, it may be that either the lowest or first radially excited state has been misidentified.  Removing either the $f_0(550)$ or the $f_0(980)$ would shift all the higher masses to the left by one unit of $n$, resulting in a much better fit to the model.  An obvious extension of this work would be to include strange quarks and glueballs and to determine the mixing among the resulting scalar states.

\begin{table}[h!]
\begin{center}
\begin{tabular}{|c||c|c|}
\hline
$n$ & $f_0$ experimental (MeV) & $f_0$ model (MeV) \\
\hline
\hline
1 & $550^{+250}_{-150}$ & $799$ \\
\hline
2 & $980 \pm 10$ & $1184$ \\
\hline
3 & $1350 \pm 150$ & $1466$ \\
\hline
4 & $1505 \pm 6$ & $1699$ \\
\hline
5 & $1724 \pm 7$ & $1903$ \\
\hline
6 & $1992 \pm 16$ & $2087$ \\
\hline
7 & $2103 \pm 8$ & $2257$ \\
\hline
8 & $2314 \pm 25$ & $2414$ \\\hline
\end{tabular}
\caption{The experimental and predicted values of the scalar meson masses.}
\label{scalarmasses}
\end{center}
\end{table}

\subsection{Vector sector} 
\label{secVector}

The soft-wall model with the dilaton $\phi(z)=\lambda z^{2}$ describes the $\rho$ meson spectrum surprisingly well \cite{Karch:2006pv}. In fact, since the scalar field VEV does not couple to the vector sector, any dilaton with the behavior (\ref{dilatonlz}) causes the vector mass spectrum to exhibit linear trajectories for the higher resonances. Examining the QCD experimental data, one sees that the $\rho$ mass spectrum exhibits linear behavior around 
$\rho(1465)$ or $\rho(1720)$; therefore, one expects the appropriately modified dilaton as $z\rightarrow 0$ will 
only affect lower lying resonances as higher eigenfunctions localize towards the IR and are less dependent on 
small $z$ behavior.  

From the action (\ref{action1}) the equation of motion of the vector field $V_\mu^n(x,z) = {\cal V}_\mu^n(x)V_n(z)$ 
using the axial gauge $V_5=0$ is given by
\begin{equation} \label{swvector}
-\partial_z^{2}V_n+\omega'\partial_z V_n=m_{V_n}^2 V_n,
\end{equation}
where $\omega=\phi(z)+\log{z}$.
With the substitution $V_{\mu}^{n}=e^{\omega/2} v_{n}$, (\ref{swvector}) can be written in the Schr\"{o}dinger form,
\begin{equation} \label{schroedinger}
-\partial_z^{2}v_{n}+\left(\frac{1}{4}\omega'^{2} - \frac{1}{2}\omega''\right) v_{n}=m_{V_n}^{2}v_{n}.
\end{equation}
Using the dilaton form $\phi = \lambda z^{2}$, the eigenvalues of (\ref{schroedinger}) at large $n$ can be solved analytically with the boundary conditions $\lim_{z_0\rightarrow 0} v_n(z_0)=0$, $ \partial_z v_{n}(z\rightarrow\infty)=0$ 
and agree with those found in \cite{Karch:2006pv}, namely
\begin{equation} \label{scaleeigenvalues}
m_{V_n}^{2}\approx\left(4n+4\right)\lambda, \quad n=0,1,2,\ldots
\end{equation}
\begin{figure}[h!]
\begin{center}
\includegraphics[scale=0.45,angle=90]{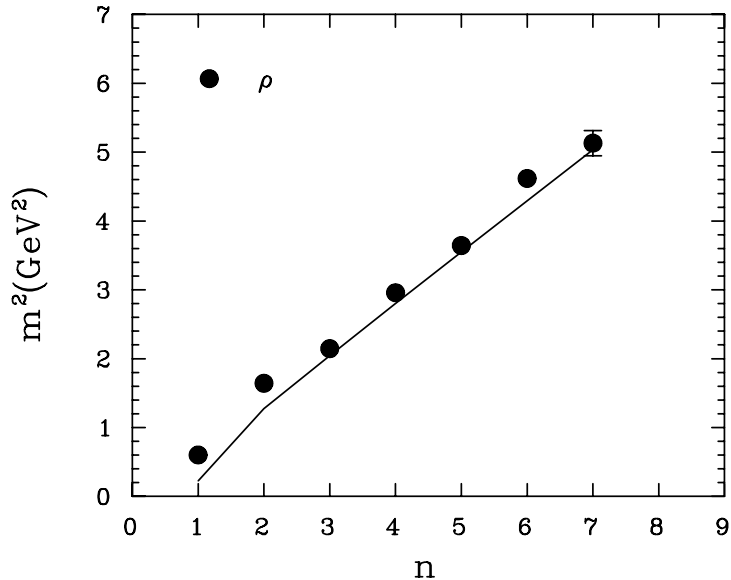}
\caption{\textsl{Comparison of the predicted vector mass eigenvalues using the tanh form (\ref{arcv}) of $v(z)$ (solid) with the QCD $\rho$ mass spectrum \cite{pdg}. 
}}
\label{VectorMasses}
\end{center}
\end{figure}

\begin{table}[h!]
\begin{center}
\begin{tabular}{|c||c|c|}
\hline
$n$ & $\rho$ experimental (MeV) & $\rho$ model (MeV) \\
\hline
\hline
1 & $775.5 \pm 1$ & $475$ \\
\hline
2 & $1282 \pm 37$ & $1129$ \\
\hline
3 & $1465 \pm 25$ & $1429$ \\
\hline
4 & $1720 \pm 20$ & $1674$ \\
\hline
5 & $1909 \pm 30$ & $1884$ \\
\hline
6 & $2149 \pm 17$ & $2072$ \\
\hline
7 & $2265 \pm 40$ & $2243$ \\
\hline
\end{tabular}
\caption{The experimental and predicted values of the vector meson masses.}
\label{vectormasses}
\end{center}
\end{table}
where $\lambda$ sets the scale for the vector meson Kaluza-Klein tower. However, since the dilaton specified in (\ref{phieqn}) is modified for $z\lesssim 1$ there is a change in the slope of the mass spectrum around $n=2$ which matches the behavior of the experimental data. The numerical vector mass spectrum is compared to the experimental data in Figure~\ref{VectorMasses} and displayed in Table \ref{vectormasses}. While the prediction for the $\rho(775)$ mass is low, the rest of the vector meson masses are in reasonable agreement with experiment.  Most likely the agreement with the $\rho(775)$ could be improved upon by using a parameterization of $v(z)$ which rises more rapidly to its asymptotic value at large $z$.  Nevertheless, the purpose of this paper is to incorporate QCD-like chiral symmetry breaking in soft-wall 
AdS/QCD models, not just to fit data.

\subsection{Axial-vector sector} 
\label{secAxial}

Unlike the vector field, the axial-vector couples to the scalar field VEV, producing a $z$-dependent mass term in its equation of motion. Similarly to the vector field case, the equation of motion assuming $A_\mu(x,z)={\cal A}_\mu^n(x) A_n(z)$ using the axial gauge $A_5=0$ is given by 
\begin{equation} \label{swaxial}
-\partial_z^{2}A_n+\omega'\partial_z A_n+g_5^2 \frac{L^2}{z^2} v^2(z) A_n=m_{A_n}^{2}A_n.
\end{equation}
Using the same transformation as for the vector field, $A_n=e^{\omega/2} a_n$, one can express (\ref{swaxial}) as
\begin{equation} \label{swaxialtrans}
-\partial_z^{2}a_{n}+\left(\frac{1}{4}\omega'^2 - \frac{1}{2}\omega'' +g_5^2 
\frac{L^2}{z^2} v^2(z) \right)a_n = m_{A_n}^2 a_n.
\end{equation}
The expression (\ref{swaxialtrans}) for the axial-vector field matches that of the vector field except for the 
additional term, $g_5^2 v^2(z) L^2/z^2$. Because of this $z$-dependent mass term, equation (\ref{swaxialtrans}) is difficult to solve analytically and again requires a numerical solution using the shooting method. Using the boundary conditions $\lim_{z_0\rightarrow 0} a_n(z_0)=0$, $ \partial_z a_{n}(z\rightarrow\infty)=0$, the axial-vector meson spectrum is obtained for the fixed values of $\lambda$, $m_q$, $\Sigma$, and $\kappa$, and match the $a_1$ experimental data \cite{pdg}. 

\begin{figure}[h!]
\begin{center}
\includegraphics[scale=0.45,angle=90]{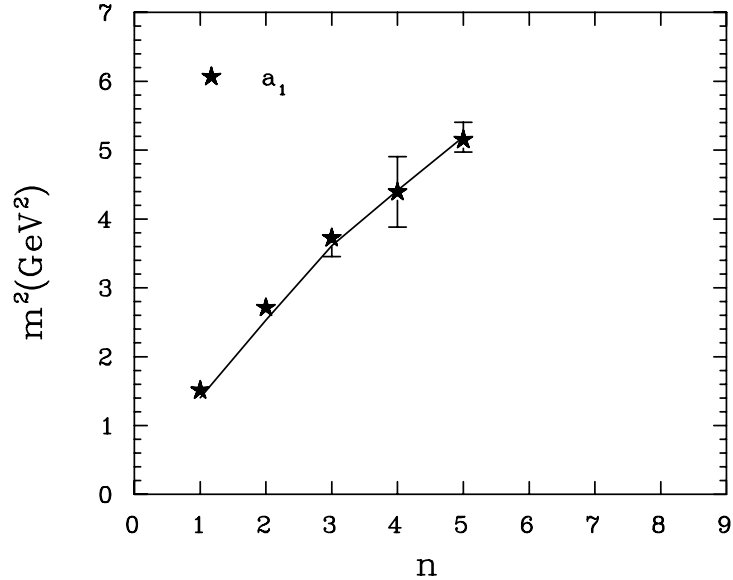}
\caption{\textsl{Comparison of the numerical results for the axial-vector mass eigenvalues using the tanh form (\ref{arcv}) of $v(z)$ (solid)
with the QCD $a_1$ mass spectrum \cite{pdg}.
}}
\label{AxialMasses}
\end{center}
\end{figure}

The limiting behavior of $v(z)$ as $z\rightarrow\infty$ leads to a constant shift between the vector and axial-vector spectra at high mass values.  Comparing the equations of motion (\ref{schroedinger}) and (\ref{swaxialtrans})
for these fields one finds the asymptotic behavior
\begin{equation}
\Delta m^2 \equiv \left(m_{A_n}^2 - m_{V_n}^2\right)_{n \rightarrow \infty}
= g_5^2 \frac{L^2}{z^2} v^2(z \rightarrow \infty) = \frac{4g_5^2 \lambda}{\kappa}.
\label{deltam2}
\end{equation}
Together with the slope $\lambda$, this determines the numerical value to be $\kappa\sim 30$, although the best visual global fit to all the data suggests $\kappa = 15$, which is the value used here.  This is probably due to the small number of radial excitations to which we are fitting.  The results of our analysis are plotted in Figure \ref{AxialMasses} and displayed in Table~\ref{avmasses}. The $a_1(1260)$ resonance is predicted to within 5$\%$ and there is good agreement with the higher resonances of $a_1$. 

Note that from (\ref{deltam2}), $\Delta m^2 >0$ implies that $\kappa >0$, which means
that the potential in (\ref{action1}) is unbounded from below.  To address the stability of the gravity-dilaton background requires a complete fluctuation analysis generalizing the work in \cite{bl}. Even though this is beyond the scope of the present work it does suggest that higher-order terms will be needed for stability.

\begin{table}[h!]
\begin{center}
\begin{tabular}{|c||c|c|}
\hline
$n$ & $a_1$ experimental (MeV) & $a_1$ model (MeV) \\
\hline
\hline
1 & $1230 \pm 40$ & $1185$ \\
\hline
2 & $1647 \pm 22$ & $1591$ \\
\hline
3 & $1930^{+30}_{-70}$ & $1900$ \\
\hline
4 & $2096 \pm 122$ & $2101$ \\
\hline
5 & $2270^{+55}_{-40}$ & $2279$ \\
\hline
\end{tabular}
\caption{The experimental and predicted values of the axial-vector meson masses.}
\label{avmasses}
\end{center}
\end{table}

\section{Pion Coupling}
\label{secPion}

To confirm that our model is in fact consistent with AdS/QCD model predictions, we calculate the pion decay constant using the formula \cite{Erlich:2005qh}
\begin{equation}
f_\pi^2=-\frac{1}{g_5^2} \lim_{\epsilon\rightarrow 0}
\frac{\partial_z A_0(0,z)}{z}\Bigg |_{z=\epsilon}
\end{equation}
where with our setup we calculate $f_\pi=92.4$ MeV. Here $A_0(q, z)$ is the axial-vector bulk-to-boundary propagator with boundary conditions $A_0(0,\epsilon)=1$ and $\partial_z A_0(0,z \rightarrow \infty)=0$, and $-q^2$ replaces $m_{A_n}^2$. Using the Gell-Mann-Oakes-Renner relation and the measured values of the pion decay constant and pion mass, the above formula returns the measured value of the pion decay constant if $m_q = 9.75$ MeV, $\kappa = 15$, and $\lambda = 0.1831$ GeV$^2$.

Although we do not solve for the mass spectra of the pseudoscalar mesons in this paper, because it involves solving a fourth order differential equation, we have calculated the ground-state pion mass and can determine its vector coupling $g_{\rho\pi\pi}$. The $V\pi\pi$ coupling is given in \cite{Kwee:2007nq} as 
\begin{equation} \label{grpp}
g_{\rho_{n}\pi\pi} =  \frac{1}{f_{\pi}^2} \int dz \,V_n(z) e^{-\phi(z)} \left(\frac{1}{g_{5} z}(\partial_{z}\varphi(z))^{2} + \frac{g_5 L^2 v^2(z)}{z^{3}}(\pi(z)-\varphi(z))^{2}\right)~,
\end{equation}
where $V_n$ are the rho-meson Kaluza-Klein wave functions. They are normalized as follows
\begin{equation}
\int dz\, \frac{e^{-\phi(z)}}{z} V_n(z) V_m(z) = \delta_{mn}.
\end{equation}
The functions $\pi(z)$ and $\varphi(z)$ must be determined from the system of equations for the axial-vector and pion as given in \cite{Kwee:2007nq}
\begin{eqnarray}
\left[ e^{\phi} \partial_{z}\left(\frac{e^{-\phi}}{z}
\partial_{z}A_{\mu}\right)-\frac{q^{2}}{z}A_{\mu}-\frac{g_{5}^{2} L^2 v^2(z)}{z^{3}}A_{\mu}\right]_{\perp}=0,\label{Aperp} \\
e^{\phi} \partial_{z}\left(\frac{e^{-\phi}}{z}\partial_{z}\varphi\right)
+\frac{g_{5}^{2} L^2 v^2(z)}{z^{3}}(\pi - \varphi)=0, \label{mixed1}\\
q^{2}\partial_{z}\varphi + \frac{g_{5}^2 L^2 v^2(z)}{z^{2}} 
\partial_{z}\pi =0,\label{mixed2}
\end{eqnarray}
where $A_{\mu} = A_{\mu\perp} + \partial_{\mu}\varphi$. The pion is then the solution to equations (\ref{mixed1}) and (\ref{mixed2}).  Following exactly the same steps as \cite{Erlich:2005qh} one may derive the Gell-Mann-Oakes-Renner relation from this set of equations.

\begin{figure}[h!]
\begin{center}
\includegraphics[scale=0.49,angle=90]{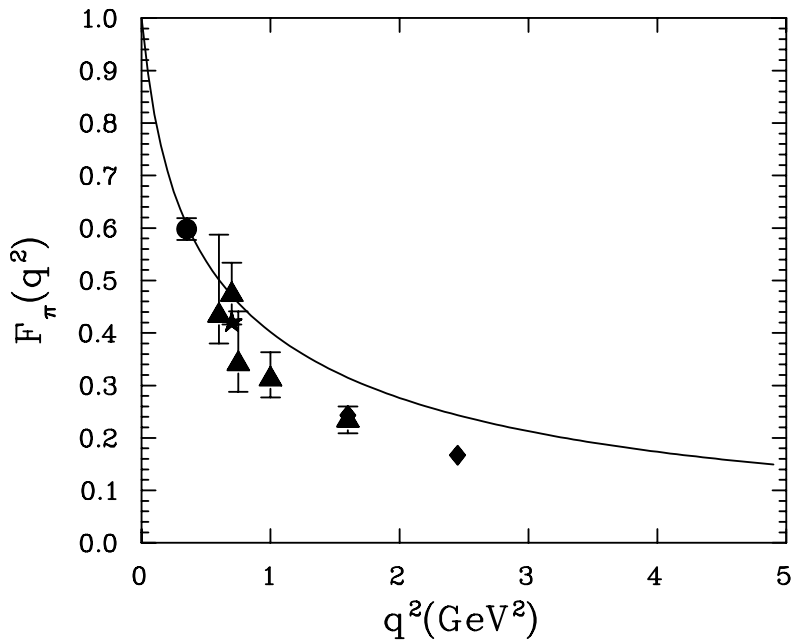}
\caption{\textsl{The line shows the predicted space-like behavior of the pion form factor $F_{\pi}(q^2)$ which is compared to the experimental data obtained from \cite{Kwee:2007nq}. The triangles are data from DESY, reanalyzed by \cite{Tadevosyan:2007yd}. The diamonds are data from Jefferson Lab \cite{Horn:2006tm}. The circles \cite{Ackermann:1977rp} as well as the star \cite{Brauel:1977ra} are also data obtained from DESY.}}
\label{Fpiplot}
\end{center}
\end{figure}

The expression for $g_{\rho_{n}\pi\pi}$ can be approximated by setting $\varphi(z) = A_{0}(0,z) - 1$ and 
$\pi(z) = -1$. Previous soft-wall models \cite{Kwee:2007nq} have obtained values smaller than the experimental result of $g_{\rho\pi\pi}\approx 6$. Similarly, our calculations also give a low value $g_{\rho\pi\pi} =2.89$.  Once we calculate the $g_{\rho_{n}\pi\pi}$, the space-like pion form factor can easily be determined from a sum over vector meson poles,
\begin{equation} \label{Fpisum}
F_{\pi}(q^{2}) = \sum_{n=1}^{\infty}\frac{f_{n}g_{\rho_{n}\pi\pi}}{q^{2}+m_{V_n}^{2}},
\end{equation} 
where $f_{n}$ are the decay constants of the vector modes. However (\ref{Fpisum}) converges slowly\,\footnote{TMK thanks Herry Kwee for correspondence on this issue.} and 
numerically it is much better to use the expression in terms of the vector and axial-vector bulk-to-boundary propagators as in \cite{Kwee:2007nq}
\begin{equation}
F_{\pi}(q^{2}) =\int{dz\, e^{-\phi(z)} \frac{V_0(q,z)}{f_{\pi}^{2}}\left(\frac{1}{g_5^2 z} (\partial_{z}\varphi(z))^{2} +\frac{v^2(z)}{z^{3}}(\pi(z)-\varphi(z))^2\right)},
\end{equation}
where $V_0(q,z)$ is the vector bulk-to-boundary propagator.
The results of our $F_{\pi}(q^{2})$ calculation are plotted in Figure \ref{Fpiplot}, and shows a slight improvement in matching the experimental values compared to that obtained in Ref.~\cite{Kwee:2007nq}. 
It is apparent that the QCD pion behavior is mimicked reasonably well, beyond that expected from the simple soft-wall AdS/QCD model.

\section{Conclusion} \label{secDiscuss}

We have shown how to incorporate chiral symmetry breaking into a soft-wall version of the AdS/QCD model
with independent sources for spontaneous and explicit breaking. This is achieved by introducing a quartic term in the potential for the bulk scalar field dual to the quark bilinear operator ${\bar q} q$. This changes the dilaton profile for small $z$, while simultaneously maintaining the large $z$ quadratic behavior and therefore linear trajectories for the radially excited states. In addition, our model is built from the assumption of preserving chiral symmetry for highly excited states, which is supported by the experimental values of the QCD mass spectrum. This enables us to obtain reasonable agreement within 10$\%$ of the QCD meson mass spectra for scalar, vector and axial-vector fields, although the lowest lying $\rho$ and $f_0$ predictions
are not as good.

Even though our modification of the soft-wall version of the AdS/QCD model is simple and predictive,
any further progress must recognize the limitations of this type of phenomenological model. Genuine
stringy behavior is most likely required to fully describe the characteristics of QCD. Nevertheless some
features such as masses and couplings seem to agree better than expected and it would be worth 
using the modified dilaton profile to study further details of the meson spectrum. On the theoretical side
it would be interesting to further understand the soft-wall model from the top-down including finding a 
dynamical solution of the features exhibited in our model along the lines considered in Ref.~\cite{Batell:2008zm}. 
In addition the stability of the scalar potential will most likely require higher-order terms that can only be 
studied from the top-down. It is interesting that the simple 5D model contains QCD-like features and suggests 
that a further understanding of QCD can be obtained from the gauge/gravity correspondence.

\section*{Acknowledgments}
TMK thanks Brian Batell and Todd Springer for insightful discussions. The work of TG and TMK was supported in part by the Research Corporation for Science Advancement. TG is also supported by the Australian Research Council. The work of JIK and TMK is supported by the US Department of Energy (DOE) under grant DE-FG02-87ER40328. TMK is also supported by a Fellowship from the School of Physics and Astronomy at the University of Minnesota.

\end{document}